\documentclass[12pt]{article}%
\usepackage{amsmath}
\usepackage{amsfonts}
\usepackage{amssymb}
\usepackage{graphicx}%
\providecommand{\U}[1]{\protect\rule{.1in}{.1in}}
\newcommand{\bfr}{\begin{flushright}}
\newcommand{\efr}{\end{flushright}}
\newcommand{\bc}{\begin{center}}
\newcommand{\ec}{\end{center}}
\newcommand{\ben}{\begin{enumerate}}
\newcommand{\een}{\end{enumerate}}

\newcommand{\be}{\begin{equation}}
\newcommand{\ee}{\end{equation}}
\newcommand{\ba}{\begin{array}}
\newcommand{\ea}{\end{array}}

\def\6{\partial}
\newcommand{\bea}{\begin{eqnarray}}
\newcommand{\eea}{\end{eqnarray}}
\begin{document}

\title{Quantization of the Relativistic Fluid in Physical Phase Space on K\"{a}hler Manifolds}
\author{L. Holender$^{a}$\thanks{email: holender@ufrrj.br}\, M. A. Santos$^{b}%
$\thanks{email: masantos@cce.ufes.br}\, and I. V. Vancea$^{a}$\thanks{email:
ionvancea@ufrrj.br}}
\date{January 21, 2008}
\maketitle

\pagestyle{empty}

\begin{center}
\emph{$^{a}$Departamento de F\'{\i}sica, Universidade Federal Rural do Rio de
Janeiro (UFRRJ),\newline Cx. Postal 23851, 23890-000 Serop\'{e}dica - RJ,
Brasil}\newline\emph{$^{b}$Departamento de F\'{\i}sica e Qu\'{\i}mica,
Universidade Federal do Esp\'{\i}rito Santo (UFES), Avenida Fernando Ferarri
S/N - Goiabeiras, 29060-900 Vit\'{o}ria - ES, Brasil}
\end{center}

\abstract{
We discuss the quantization of a class of relativistic fluid models defined in terms of one real and two complex
conjugate potentials with values on a K\"{a}hler manifold, and parametrized by the K\"{a}hler potential
$K(z,\overline{z})$ and a real number $\lambda$. In the hamiltonian formulation, the canonical conjugate
momenta of the potentials are subjected to second class constraints which allow us to apply
the symplectic projector method in order to find the physical degrees of freedom and the physical hamiltonian.
We construct the quantum theory for that class of models by employing the
canonical quantization methods. We also show that a semiclassical theory in which the K\"{a}hler and the
complex potential are not quantized has a highly degenerate vacuum. Also, we define and compute the
quantum topological number (quantum linking number) operator which has non-vanishing contributions from the K\"{a}hler and complex potentials
only.
Finally, we show that the vacuum and the states formed by tensoring the number operators eigenstates have zero
linking number and show that linear combinations of the tensored number operators eigenstates which have the
form of entangled states have non-zero linking number.}

\newpage

\section{Introduction}

The recent interest in the lagrangian and hamiltonian formulation of the
relativistic fluid mechanics with non-zero vorticity has been motivated by a
large range of applications such as: the neutron star dynamics \cite{sja}, the
representation of the massless axion field in cosmology \cite{vv}, the
Kalb-Ramond and global string induced superfluid vorticity \cite{cl,ds}, and
the relationship between the superstrings, supermembranes and the superfluid
dynamics \cite{jpp1,jpp2,rj1,rj2}.

The analysis of the vorticity in the lagrangian formalism requires a
particular parametrization of the fluid potentials. In the case of the
non-relativistic fluid, this is achieved by the Clebsch parametrization of the
velocity potential $\mathbf{v}=\mathbf{\nabla}\theta+\alpha\mathbf{\nabla
}\beta$, where $\alpha$, $\beta$ (the Gauss potentials) and $\theta$ are
scalar fields \cite{ac}. In the hamiltonian formulation, the Clebsch
parametrization allows to express the kinematic helicity as a surface integral
and to realize canonically the non-vanishing brackets between the variables
$\rho$ (matter density field) and $\mathbf{v}$ \cite{jpi}. In \cite{jpp2}, it
was proposed a different parametrization of the fluid potentials in which the
vortex dynamics is expressed in terms of grassmannian coordinates and the
vortex is associated to the spin. That parametrization has been used for the
generalization of the Chaplygin gas model in $d=2+1$ and $d=3+1$ \cite{sc}(see
for an introduction and reviews \cite{jlect,bc}.)

Another very interesting proposal for a parametrization of the fluid
potentials was put forward in \cite{nhg1}, where the potentials are complex
coordinates on an arbitrary K\"{a}hler manifold. That approach allows to
construct interesting models of supersymmetric fluids which, on their turn,
shade some light on the problem of the superfluid phases of certain
supersymmetric systems \cite{bj,mh,nhg}. The parametrization of the potentials
in terms of complex coordinates preserves the property of the Clebsch
parametrization, namely to elliminate the obstruction to building a consistent
lagrangian which could exist due to a Chern-Simons term that is necessary in
order to consider the non-zero vorticity \cite{jlect}. The complex
parametrization can be introduced instead of the Clebsch parametrization
already at the non-supersymmetric level and the full structure of the
divergence free currents and the topological charges associated to non-zero
vorticity can be obtained in this way \cite{nhg1}. An important feature of the
complex parametrization is that the hamiltonian dynamics is governed by a set
of simple second-class constraints among the fluid degrees of freedom. In this
paper, we propose a quantized theory for a large set of non-supersymmetric
relativistic fluid models on K\"{a}hler manifolds.

The fluid properties of the system under consideration are described by a set
of fields that represent the fluid potentials and which, together with their
canonically conjugate momenta, can be organized to form a canonical phase
space. The potentials and their momenta are subjected to second class
constraints which form a set of simple algebraic equations. As a first result,
we will obtain the physical degrees of freedom of the largest set of classical
fluid models by applying the symplectic projector method developped in
\cite{amar1,amar2,sjp} and more recently employed in the study of the extended
abelian Chern-Simons theory \cite{mnm}, the noncommutative open string
\cite{mmi}, the Maxwell-Chern-Simons theory \cite{hmi}, the Lorentz-symmetry
violation \cite{bmm}, quantum gravity potentials \cite{rc} and cosmological
perturbations \cite{db} (see for reviews \cite{mmi-r1,mmi-r2}.) Althought
equivalent to other methods used to solve the second class constraints, the
symplectic projector method has the advantage of giving a geometrical picture
to the relationship between the set of fields and their canonical conjugate
momenta on one hand and the set of physical degrees of freedom defined by
local coordinates on the constraint surface on the other hand. Compared with
the literature, our results reproduce exactly the general classical
hamiltonian from \cite{nhg1}. Moreover, there are some new results consisting
in the explicit identification of the physical degrees of freedom and of their
relationship with the linearly dependent fields. The main result of this paper
is the quantization of a smaller class of models parametrized by the
K\"{a}hler potential $K(z,\overline{z})$ and a real parameter $\lambda$. That
corresponds to choosing the arbitrary potential function on the fluid density
$f(\rho)$ of the form $f(\rho)=\lambda\rho^{2}/2$ by which a set of
quantizable fluid models is selected. The larger class of classical models is
parametrized by $K(z,\overline{z})$ and the general potential functions
$f(\rho)$ and, in general, it is not suitable for quantization due to the
arbitrary form each of these two functions can take. Our approach to the
quantum fluid is based on the canonical quantization performed in the physical
phase space of the relativistic fluid. However, since the K\"{a}hler and the
complex potentials determine a $d=3$ vector field $\mathbf{A}(z,\overline{z})$
associated to the conserved currents, we quantize the field $\mathbf{A}$
rather than the complex fields $z$ and $\overline{z}$. Due to the large
arbitrariness in choosing the K\"{a}hler potential $K(z,\overline{z}),$ there
is no simple relationship between the quantized fields $\mathbf{A}$ and $z$
and $\overline{z}$. That suggest that one could also consider a semiclassical
situation in which the potential $\mathbf{A}$ is kept classical and just the
canonical pair $\theta$ and $\pi_{\theta}$ is quantized. However, it turns out
that in this case the Fock space is highly degenerate having all one-particle
states proportional to the vacuum. Therefore, the quantization of the complex
potentials is imposed, and quantizing the field $\mathbf{A}$ represents the
simplest approach to this problem. The last new result of this paper is the
calculation of the quantum topological charge (or quantum linking number)
operator $\widehat{\omega}$. This is constructed from the normal ordered
expression of the classical topological charge. The normal ordering is
necessary because there are terms in $\widehat{\omega}$ which have the general
form $\delta^{3}\left(  0\right)  k^{n}$, where $k$ is any component of the
momentum and $n=1,2,3$. These terms are undetermined in the low momentum limit
and vanish otherwise. We discuss the structure of the Fock space with respect
to $\widehat{\omega}$ and show that the vacuum and the number operator
eigenstates have zero linking number while the linear combination of tensored
number operators eigenstates have non-zero topological charge \footnote{The
quantum fluid states with non-zero linking number are actually entangled
number operators eigenstates.}.

The paper is organized as follows. In Section 2 we briefly review the general
relativistic fluid on Kh\"{a}ler manifolds in the hamiltonian formulation from
\cite{nhg1}. Also, we determine the physical degrees of freedom which are the
local coordinates on the constraint manifold by the symplectic projector
method and write down the expression of the physical hamiltonian and the
conserved charges in term of these. The physical hamiltonian obtained by
applying the symplectic method coincides with the one given in the literature.
In Section 3 we quantize the particular class of models parametrized by
$K(z,\overline{z})$ and $\lambda$. We give the structure of the Fock space and
show that if the vector potential $\mathbf{A}$ is considered as a classical
field, the Fock space of the semiclassical quantized fluid is highly
degenerate. In Section 4 we discuss the quantum topological operator and
determine the general form of the states from the Fock space which have
non-vanishing linking number. The last section is devoted to conclussions.

\section{Relativistic Fluid in the Hamiltonian Formulation}

In this section we are going to review the non-supersymmetric perfect
relativistic fluid studied in \cite{nhg1}. Also, we are going to derive the
physical degrees of freedom in a new way by applying the symplectic projector
method developped in \cite{mmi-r1,mmi-r2}. That is equivalent, of course, to
solving the constraint equations in a purely algebraic way as was done in
\cite{nhg1}. However, the symplectic projector method provides a simple
geometrical picture for the relationship among potentials, physical degrees of
freedom and constraint surface.

In the class of models under consideration, the relativistic fluid is
characterized by the equations of state for the local physical quantities $p$,
$\varepsilon$, and $\rho$ which are the pressure, the energy density and the
local fluid density, respectively. The dynamics conserves the energy-momentum
tensor $T_{\mu\nu}$ and the fluid density current $j^{\mu}$%
\begin{equation}
\partial^{\mu}T_{\mu\nu}=0,\qquad\partial_{\mu}j^{\mu}=0,\label{1.1}%
\end{equation}
where
\begin{equation}
T_{\mu\nu}=p\eta_{\mu\nu}+(\varepsilon+p)u_{\mu}u_{\nu},\qquad j^{\mu}=\rho
u^{\mu}.\label{1.2}%
\end{equation}
Here, $\eta_{\mu\nu}=(-,+,+,+)$ is the Minkowski metric, $u^{\mu}=dx^{\mu
}/d\tau$ is the velocity four-vector with $u_{\mu}^{2}=-1$ and $\tau$ is the
proper time along the flow line of the current. From the current conservation
given in the equations (\ref{1.1}), one can see that there are three
independent current components to which one can assign three fluid potentials
$(\theta,z,\bar{z})$ that play the role of the Lagrange multipliers in the
lagrangian formulation of the theory. As was argued in \cite{nhg1}, one can
take $\theta$ real, $z$ complex and $\bar{z}$ the complex conjugate of $z$.
One can choose the complex potentials to parametrize an arbitrary K\"{a}hler
manifold characterized by the K\"{a}hler potential $K(z,\bar{z})$ which is a
real function on $z$ and $\bar{z}$. The above conservation equations can be
obtained from the following lagrangian density \cite{nhg1}
\begin{equation}
\mathcal{L}[j^{\mu},\theta,\bar{z},z]=-j^{\mu}\left(  \partial_{\mu}%
\theta+i\partial K\partial_{\mu}z-i\overline{\partial}K\partial_{\mu}%
\overline{z}\right)  -f(\rho),\label{1.8}%
\end{equation}
where $\partial K=\partial_{z}K$, $\overline{\partial}K=\partial_{\overline
{z}}K$ and $f(\rho)$ is some potential function on $\rho$. Also, one can see
from the definition of the fluid density current that $\rho=\sqrt{-j^{2}}$.
The action is invariant under the spacetime translations and the potential
fields reparametrizations. The corresponding conservation laws are: the
conservation of the energy-momentum tensor and the conservation of the fluid
density current as given in the equations (\ref{1.1}), and the conservation of
an infinity of reparametrization currents $J_{\mu}[G]=-2G(\bar{z},z)j_{\mu}$,
where $G(\bar{z},z)$ are arbitrary analytic functions on $z$ and $\overline
{z}$. Beside the reparametrization currents given above, there are conserved
axial currents generated by conserved topolgical charges defined by the
following relation \cite{nhg1}%
\begin{equation}
\omega=-2i\int d^{3}x\,\partial_{i}\left[  \varepsilon^{ijk}\theta
\partial\overline{\partial}K\partial_{j}\bar{z}\,\partial_{k}z\right]
.\label{classical-topological-number}%
\end{equation}
The charges $\omega$ can be interpreted as the linking number of vertices
formed in the fluid.

The fluid flow and the non-zero vorticity can be described in the hamiltonian
formalism, too. Following \cite{nhg1}, we define the canonically conjugate
momenta as follows
\begin{align}
\pi_{\mu}  &  =\left.  \frac{\partial\mathcal{L}}{\partial u^{\mu}}\right\vert
_{\rho}=\rho\left(  \partial_{\mu}\theta+i\partial K\partial_{\mu}%
z-i\overline{\partial}K\partial_{\mu}\bar{z}\right)  ,\qquad\pi_{\theta}%
=\frac{\partial{\mathcal{L}}}{\partial\partial_{0}{\theta}}=j_{0}%
,\label{2.11a}\\
\pi_{z}  &  =\frac{\partial{\mathcal{L}}}{\partial\partial_{0}{z}}=i\partial
Kj_{0},\qquad\pi_{z}=\frac{\partial{\mathcal{L}}}{\partial\partial
_{0}\overline{z}}=-i\overline{\partial}Kj_{0}. \label{2.11b}%
\end{align}
Note that the currents ${j_{\mu}}$ do not appear dynamically in the theory and
that the axial current depends locally on $\pi_{\mu}$. Therefore, the relevant
phase space for the physical degrees of freedom is the reduced phase space
$\left(  \theta,z,\bar{z},\pi_{\theta},\pi_{z},\pi_{z}\right)  $. The
equations (\ref{2.11b}) represent a set of two second class constraints in the
reduced phase space
\begin{equation}
\Omega_{1}=\pi_{z}-i\partial K\pi_{\theta}=0,\hspace{2em}\Omega_{2}=\pi
_{\bar{z}}+i\overline{\partial}K\pi_{\theta}=0. \label{2.13}%
\end{equation}

The physical degrees of freedom of the relativistic fluid can be obtained from
the reduced phase space potentials by applying the symplectic projector method
by which the reduced phase space is projected on to the constraint surface
defined by the relations (\ref{2.13}) \cite{sjp,mmi-r1,mmi-r2}. Let us
introduce the following notation for the potentials and their momenta
\begin{equation}
\{\xi_{i}\}=\left\{  \xi_{1,}\xi_{2,}\xi_{3,}\xi_{4,}\xi_{5,}\xi_{6}\right\}
=\left\{  \theta,z,\bar{z},\pi_{\theta},\pi_{z},\pi_{z}\right\}
.\label{field-redefinition}%
\end{equation}
The local coordinates on the constraint surface $\{\xi_{i}^{\ast}\}$ are
obtained by applying the symplectic projector $\Lambda$ to the reduced phase
space fields $\{\xi_{i}\}$%
\begin{equation}
\xi_{i}^{\ast}(x^{0},\mathbf{x})=\int d^{3}\mathbf{y}\sum_{j=1}^{6}\Lambda
_{i}^{j}(x^{0},\mathbf{x},\mathbf{y})\xi_{j}(x^{0},\mathbf{y}),
\end{equation}
where the general form of the symplectic projector is given by the following
relation
\begin{equation}
\Lambda_{j}^{i}(x^{0},\mathbf{x},\mathbf{y})=\delta_{j}^{i}\delta^{3}\left(
\mathbf{x}-\mathbf{y}\right)  -J^{ik}\displaystyle%
{\displaystyle\int}
d^{3}\mathbf{z}d^{3}\mathbf{w}\frac{\delta\Omega_{\alpha}(x^{0},\mathbf{z}%
)}{\delta\xi_{k}(x^{0},\mathbf{x})}D_{\alpha\beta}^{-1}(x^{0},\mathbf{z}%
,\mathbf{w})\frac{\delta\Omega_{\beta}(x^{0},\mathbf{w})}{\delta\xi_{j}%
(x^{0},\mathbf{y})}.\label{L-projector}%
\end{equation}
Here $J^{ik}=-J^{ki}$ is the symplectic matrix of the reduced phase space and
$D_{\alpha\beta}^{-1}$, $\alpha,\beta=1,2$, is the inverse of the Dirac matrix
of the constraint brackets computed at equal times. It is a simple exercise to
show that the symplectic projector has the following expression
\begin{equation}
\Lambda(x^{0},\mathbf{x},\mathbf{y})=\left(
\begin{array}
[c]{cccccc}%
1 & 0 & 0 & 0 & \frac{\overline{\partial}K}{2\partial\overline{\partial}%
K\xi_{4}} & \frac{\partial K}{2\partial\overline{\partial}K\xi_{4}}\\
0 & \frac{1}{2} & 0 & -\frac{\overline{\partial}K}{2\partial\overline
{\partial}K\xi_{4}} & 0 & \frac{i}{2\partial\overline{\partial}K\xi_{4}}\\
0 & 0 & \frac{1}{2} & -\frac{\partial K}{2\partial\overline{\partial}K\xi_{4}}
& -\frac{i}{2\partial\overline{\partial}K\xi_{4}} & 0\\
0 & 0 & 0 & 1 & 0 & 0\\
0 & 0 & 0 & 0 & \frac{1}{2} & 0\\
0 & 0 & 0 & 0 & 0 & \frac{1}{2}%
\end{array}
\right)  \delta^{3}(\mathbf{x}-\mathbf{y}).\label{3.1}%
\end{equation}
From the above relation, one concludes that the number of local coordinates
$\{\xi_{i}^{\ast}\}$ on the constraint surface is equal to the number of
unprojected fields $\{\xi_{i}\}$. However, since now the system is on the
constraint surface, one can use the constraints $\{\Omega_{\alpha}\}$ to
express the coordinates $\xi_{5}^{\ast}$ and $\xi_{6}^{\ast}$ as functions of
$\xi_{4}^{\ast}$. This lefts us with the correct number of physical degrees of
freedom, that is six real degrees of freedom $\xi^{\ast}$'s which have been
obtained from the ten real fields $\xi$'s acted upon by two complex constraint
equations $\Omega_{\alpha}$. The physical hamiltonian density is given in
terms of the linearly independent fields from $\{\xi_{i}^{\ast}\}$
\begin{equation}
\mathcal{H}^{\ast}=\,\mathbf{j\cdot}\left(  \mbox{\boldmath{$\nabla$}}\xi
_{1}^{\ast}+i\partial K\mbox{\boldmath{$\nabla$}}\xi_{2}^{\ast}-i\overline
{\partial}K\mbox{\boldmath{$\nabla$}}\xi_{3}^{\ast}\right)  +f\left(
\sqrt{\left(  \xi_{4}^{\ast}\right)  ^{2}-\mathbf{j}^{2}}\right)
.\label{hamiltonian}%
\end{equation}
By using the field redefinition (\ref{field-redefinition}) one can easily see
that the expression (\ref{hamiltonian}) is exactly the hamiltonian obtained in
\cite{nhg1} by a different method. Also note that the currents $J^{\mu}[G]$
generate now the reparametrization symmetry of the physical subspace.

\section{Quantization of the Relativistic Fluid}

The theory presented in the previous section describes a large class of
relativistic fluid models parametrized by the K\"{a}hler potential
$K(z,\overline{z})$ and the function $f\left(  \sqrt{\left(  \xi_{4}^{\ast
}\right)  ^{2}-\mathbf{j}^{2}}\right)  $, respectively. Both of these
functions can have a quite general form which makes addressing the issue of
quantization difficult, if not impossible, in the general case. A less general
but still interesting class of relativistic fluid models can be obtained by
choosing a simple form for $f(\rho)$ while leaving the K\"{a}hler manifold
arbitrary. The equations of state of the reduced set of models have the
following form \cite{nhg1}%
\begin{equation}
p=\varepsilon=\frac{\lambda}{2}\,\rho^{2}~~,~~f(\rho)=\frac{\lambda}{2}%
\,\rho^{2}.\label{toy-model}%
\end{equation}
These models are parametrized by the K\"{a}hler potential $K(z,\overline{z})$
and the real number $\lambda$. From the equations of motion for $j_{l}$,
$l=1,2,3$, one can determine the explicit form of these currents in terms of
the reduced phase space fields as follows%
\begin{equation}
\frac{\partial\mathcal{H}^{\ast}}{\partial j_{l}}=0\Longrightarrow j_{l}%
=\frac{1}{\lambda}\,\left(  \partial_{l}{\xi_{1}^{\ast}}+\frac{i}{2}\partial
K\partial_{l}{\xi_{2}^{\ast}}-\frac{i}{2}\overline{\partial}K\partial_{l}%
{\xi_{3}^{\ast}}\right)  .\label{toy-model-currents}%
\end{equation}
By applying the symplectic projector formalism employed in the previous
section and by using the equations (\ref{toy-model-currents}), the following
physical hamiltonian is obtained
\begin{equation}
\mathcal{H}^{\ast}=\frac{1}{2\lambda}\left(  \mbox{\boldmath{$\nabla$}}{\xi
_{1}^{\ast}}+\frac{i}{2}\partial K\mbox{\boldmath{$\nabla$}}{\xi_{2}^{\ast}%
}-\frac{i}{2}\overline{\partial}K\mbox{\boldmath{$\nabla$}}{\xi_{3}^{\ast}%
}\right)  ^{2}+\frac{\lambda}{2}({\xi_{4}^{\ast})}^{2}.\label{mod-hamilton}%
\end{equation}
Note that $\mathcal{H}^{\ast}$  is identical to the hamiltonian calculated in
\cite{nhg1} for the models parametrized by $\{K(z,\overline{z}),\lambda\}$.

The quantization of the relativistic fluid described by the hamiltonian
(\ref{mod-hamilton}) can be performed in the canonical approach. Let us return
to the original field notations $\{\theta,\pi_{\theta},z,\overline{z}\}$
where the fields are from the physical phase subspace. The canonical conjugate
variables are $\theta$ and $\pi_{\theta}$, while $z$ and $\overline{z}$ do not
propagate. Let us introduce the following real vector potential field
\begin{equation}
\mathbf{A}(K,z,\overline{z})\equiv\mathbf{A}(x)=\frac{i}{2}\partial
K\mbox{\boldmath{$\nabla$}}z-\frac{i}{2}\overline{\partial}%
K\mbox{\boldmath{$\nabla$}}\overline{z}.\label{A-field}%
\end{equation}
From the Hamilton equations, we obtain the following equation of motion for
the $\theta$ potential
\begin{equation}
\left(  \partial^{0}\partial_{0}+\mbox{\boldmath{$\nabla$}}^{2}\right)
\theta(x^{0},\mathbf{x})=-\mbox{\boldmath{$\nabla$}}\cdot\mathbf{A}%
(x^{0},\mathbf{x}).\label{theta-eq}%
\end{equation}
This equation shows that the physical phase subspace of the relativistic fluid
is equivalent to that of a massless scalar field $\theta$ moving inside the
potential created by $K(z,\overline{z})$ and the fluid potentials $z$ and
$\overline{z}$. Actually, by assuming that $\left\vert \theta(x^{0}%
,\mathbf{x})\mathbf{A}(x^{0},\mathbf{x})\right\vert \rightarrow0$ as
$\left\vert x^{0}\right\vert \rightarrow\infty$, one can write the hamiltonian
(\ref{mod-hamilton}) as
\begin{equation}
\mathcal{H}^{\ast}=\int dx^{0}d\mathbf{x}\left[  \frac{\lambda}{2}\pi_{\theta
}^{2}+\frac{1}{2\lambda}\left(  \mbox{\boldmath{$\nabla$}}\theta\right)
^{2}+V(\theta,\mathbf{A})\right]  ,\label{hamilton-potential}%
\end{equation}
where
\begin{equation}
V(\theta,\mathbf{A})=-\frac{1}{\lambda}\left[  \theta
\mbox{\boldmath{$\nabla$}}\cdot\mathbf{A}-\frac{1}{2}\left(  \mathbf{A}%
\right)  ^{2}\right]  .\label{V-potential}%
\end{equation}
For $\mbox{\boldmath{$\nabla$}}\cdot\mathbf{A}\neq0$, the function $V$ has a
zero for each field configuration that satisfies
\begin{equation}
\theta_{0}=\frac{1}{2}\frac{\left(  \mathbf{A}\right)  ^{2}}%
{\mbox{\boldmath{$\nabla$}}\cdot\mathbf{A}}.\label{zero}%
\end{equation}
At the points from the physical phase subspace where the equations
(\ref{zero}) is satisfied, the relativistic fluid is described by the scalar
potential $\theta$ only, and it is equivalent to a free massless scalar field.
Actually, the vanishing of the gradient of the vector potential signals an
extremum of $V$. The extrema of $V$ in the $\theta$ direction and in the
$\mathbf{A}$ directions, respectively, are given in terms of physical degrees
of freedom by the following equations
\begin{equation}
\partial_{j}\left(  \partial K\partial_{j}z-\overline{\partial}K\partial
_{j}\overline{z}\right)  =0,\qquad i\left(  \partial K\partial_{j}%
z-\overline{\partial}K\partial_{j}\overline{z}\right)  =2\partial_{j}%
\theta.\label{min-eq2}%
\end{equation}
The potential $V$ takes at these points the following values
\begin{equation}
V_{1}=\frac{1}{2\lambda}\left(  \mathbf{A}\right)  ^{2}~~,~~V_{2}=-\frac
{1}{2\lambda}\left(  \theta\partial_{j}\partial^{j}\theta-\partial_{j}%
\theta\partial^{j}\theta\right)  .\label{extremum-V}%
\end{equation}
If the extremum value is obtained in all directions of physical phase subspace
simultaneously, then by equating $V_{1}$ and $V_{2}$ we obtain the following
relationship among the fluid potentials
\begin{equation}
\left(  \partial K\right)  ^{2}\mathbf{\nabla}z\cdot\mathbf{\nabla}z+\left(
\overline{\partial}K\right)  ^{2}\mathbf{\nabla}\overline{z}\cdot
\mathbf{\nabla}\overline{z}-2\partial K\overline{\partial}K\mathbf{\nabla
}z\cdot\mathbf{\nabla}\overline{z}+2\theta\left(  \mathbf{\nabla}\right)
^{2}\theta-2\mathbf{\nabla}\theta\cdot\mathbf{\nabla}\theta
=0.\label{minimum-relation}%
\end{equation}
The above equation describes the fluid configurations for which the
contribution to the fluid energy from the K\"{a}hler and the complex fluid
potentials and from their interaction with the scalar fluid potential is extreme.

For the classical models described above, the reparametrization charges $Q[G]$
have a simple form in terms of physical phase space coordinates
\begin{equation}
Q[G(\xi^{\ast})]=\int d^{3}xG(z,\overline{z})\pi_{\theta}. \label{Q-phys}%
\end{equation}
Therefore, it is easy to show that the charges $Q[G]$ are conserved provided
that the following relation holds
\begin{equation}
\int_{\Sigma}d\mathbf{s}\cdot\left(  \mathbf{\nabla}\theta+\mathbf{A}\right)
=0, \label{sp-inf-cond}%
\end{equation}
where $d\mathbf{s}$ is the area element of $\Sigma$ which is a spacelike
surface at the spatial infinity.

Since the theory is formulated in the hamiltonian formalism, it is possible to
study the quantum fluctuation of the relativistic fluid in the canonical
quantization. To this end, we interpret $\theta$ and $\mathbf{A}$ as field
operators with the dynamics given by the equation (\ref{theta-eq}) and replace
the Poisson brackets on the constraint surface by the corresponding
commutators. In what follows we are going to use the symbol $\widehat{~}$ to
denote the operators. Then the equation of motion of the quantum field
$\widehat{\theta}$ is
\begin{equation}
\left(  \partial^{0}\partial_{0}+\mbox{\boldmath{$\nabla$}}^{2}\right)
\widehat{\theta}(x^{0},\mathbf{x})=-\mbox{\boldmath{$\nabla$}}\cdot
\widehat{\mathbf{A}}(x^{0},\mathbf{x}).\label{quantum-eq-motion}%
\end{equation}
The operators $\widehat{\theta}$ and $\widehat{\mathbf{A}}$ can be decomposed
in terms of plane waves in the usual fashion
\begin{align}
\widehat{\theta}(x^{0},\mathbf{x}) &  =\int d^{3}kN_{\mathbf{k}}^{\theta
}e^{i\mathbf{k}\cdot\mathbf{x}}\widehat{a}_{\mathbf{k}}(x^{0}%
),\label{expanssion-theta}\\
\widehat{\mathbf{A}}(x^{0},\mathbf{x}) &  =\int d^{3}kN_{\mathbf{k}%
}^{\mathbf{A}}e^{i\mathbf{k}\cdot\mathbf{x}}\widehat{\mathbf{b}}_{\mathbf{k}%
}(x^{0}).\label{expanssion-A}%
\end{align}
Since the equation of motion of $\widehat{\theta}$ in the absence of the
vector potential $\widehat{\mathbf{A}}$ is the same as for a free massless
scalar field, we consider that $\widehat{\theta}$ is massless. Also, since
$\widehat{\mathbf{A}}$ is constructed from non-propagating fields
\footnote{The field $\mathbf{A}(K,z,\overline{z})$ has rather a geometric
character since it constains the information about the K\"{a}hler space
parametrized by the complex potentials $z$ and $\overline{z}$.}, we consider
the simplest situation where $\mathbf{A}$ is a massless field, too. Then it is
possible to set the normalization constants $N_{\mathbf{k}}^{\theta
}=N_{\mathbf{k}}^{\mathbf{A}}=N_{\mathbf{k}}$. By plugging the relations
(\ref{expanssion-theta}) and (\ref{expanssion-A}) into
(\ref{quantum-eq-motion}), we obtain the following set of equations for the
operators $\widehat{a}_{\mathbf{k}}(x^{0})$ and $\widehat{\mathbf{b}%
}_{\mathbf{k}}(x^{0})=\left\{  \widehat{b}_{n\mathbf{k}}(x^{0})\right\}
,n=1,2,3,$%
\begin{equation}
\left(  \partial_{0}^{2}+\mathbf{k}^{2}\right)  \widehat{a}_{\mathbf{k}}%
(x^{0})=i\mathbf{k}\cdot\widehat{\mathbf{b}}_{\mathbf{k}}(x^{0}%
),\label{eq-mot-a}%
\end{equation}
for all $\mathbf{k}$. The general solution of the equation (\ref{eq-mot-a})
can be written as
\begin{align}
\widehat{a}_{\mathbf{k}}(x^{0}) &  =\widehat{a}_{\mathbf{k}}^{(1)}%
(x^{0})e^{-\frac{i\omega_{\mathbf{k}}}{c}x^{0}}+\widehat{a}_{\mathbf{k}}%
^{(2)}(x^{0})e^{\frac{i\omega_{\mathbf{k}}}{c}x^{0}},\label{sol-a}\\
\widehat{\mathbf{b}}_{\mathbf{k}}(x^{0}) &  =\widehat{\mathbf{b}}_{\mathbf{k}%
}^{(1)}(x^{0})e^{-\frac{i\omega_{\mathbf{k}}}{c}x^{0}}+\widehat{\mathbf{b}%
}_{\mathbf{k}}^{(2)}(x^{0})e^{\frac{i\omega_{\mathbf{k}}}{c}x^{0}%
}.\label{sol-alpha}%
\end{align}
Since the classical fluid potential are real functions, i. e. $\theta
=\overline{\theta}$ and $\mathbf{A}=\overline{\mathbf{A}}$, it follows that
the corresponding quantum fields are hermitian. Therefore, the following
relations hold
\begin{equation}
\left(  \widehat{a}_{\mathbf{k}}^{(1)}\right)  ^{\dag}=\widehat{a}%
_{-\mathbf{k}}^{(2)},\left(  \widehat{\mathbf{b}}_{\mathbf{k}}^{(1)}\right)
^{\dag}=\widehat{\mathbf{b}}_{-\mathbf{k}}^{(2)}.\label{hermiticity}%
\end{equation}
By using the hermiticity condition (\ref{hermiticity}), one can write the
final form of the plane wave expanssion for the physical field operators
\begin{align}
\widehat{\theta}(x^{0},\mathbf{x}) &  =\int d^{3}kN_{\mathbf{k}}\left[
\widehat{a}_{\mathbf{k}}e^{i\left(  \mathbf{k}\cdot\mathbf{x}-\frac
{\omega_{\mathbf{k}}}{c}x^{0}\right)  }+\widehat{a}_{\mathbf{k}}^{\dag
}e^{-i\left(  \mathbf{k}\cdot\mathbf{x}-\frac{\omega_{\mathbf{k}}}{c}%
x^{0}\right)  }\right]  ,\label{theta-fin}\\
\widehat{\pi}_{\theta}(x^{0},\mathbf{x}) &  =-\frac{i}{\lambda c}\int
d^{3}kN_{\mathbf{k}}\left[  \widehat{a}_{\mathbf{k}}e^{i\left(  \mathbf{k}%
\cdot\mathbf{x}-\frac{\omega_{\mathbf{k}}}{c}x^{0}\right)  }-\widehat
{a}_{\mathbf{k}}^{\dag}e^{-i\left(  \mathbf{k}\cdot\mathbf{x}-\frac
{\omega_{\mathbf{k}}}{c}x^{0}\right)  }\right]  ,\label{mom-theta-fin}\\
\widehat{A}_{n}(x^{0},\mathbf{x}) &  =\int d^{3}kN_{\mathbf{k}}\left[
\widehat{b}_{n\mathbf{k}}e^{i\left(  \mathbf{k}\cdot\mathbf{x}-\frac
{\omega_{\mathbf{k}}}{c}x^{0}\right)  }+\widehat{b}_{n\mathbf{k}}^{\dag
}e^{-i\left(  \mathbf{k}\cdot\mathbf{x}-\frac{\omega_{\mathbf{k}}}{c}%
x^{0}\right)  }\right]  ,\label{A-fin}%
\end{align}
where $\widehat{A}_{n\mathbf{k}}$ is the $n$-th component of $\widehat
{\mathbf{A}}$\textbf{. }The normalization constant $N_{\mathbf{k}}$ can be
determined by postulating the canonical equal-time commutators
\begin{equation}
\left[  \widehat{\theta}(x^{0},\mathbf{x}),\widehat{\pi}_{\theta}%
(x^{0},\mathbf{x}^{\prime})\right]  =i\hbar\delta^{3}(\mathbf{x-x}^{\prime
}),\label{eq-time-comm}%
\end{equation}
and by defining the usual commutators among the creation and annihilation
operators
\begin{align}
\left[  \widehat{a}_{\mathbf{k}},\widehat{a}_{\mathbf{k}^{\prime}}^{\dag
}\right]   &  =\delta^{3}(\mathbf{k-k}^{\prime}),\left[  \widehat
{b}_{n\mathbf{k}},\widehat{b}_{m\mathbf{k}^{\prime}}^{\dag}\right]
=\delta_{nm}\delta^{3}(\mathbf{k-k}^{\prime})\label{comm-1}\\
\left[  \widehat{a}_{\mathbf{k}},\widehat{b}_{n\mathbf{k}^{\prime}}\right]
&  =\left[  \widehat{a}_{\mathbf{k}},\widehat{a}_{\mathbf{k}^{\prime}}\right]
=\left[  \widehat{b}_{n\mathbf{k}},\widehat{b}_{m\mathbf{k}^{\prime}}\right]
=0.\label{comm-2}%
\end{align}
By using the relations (\ref{eq-time-comm}), (\ref{comm-1}) and (\ref{comm-2}%
), one can show that the normalization constant has the following form
\begin{equation}
N_{\mathbf{k}}=\left[  \frac{\lambda\hbar c}{2\omega_{\mathbf{k}}\left(
2\pi\right)  ^{3}}\right]  ^{\frac{1}{2}}.\label{norm-constant}%
\end{equation}

The Fock space states can be constructed from the vacuum state that is
annihilated to zero by all annihilation operators in the known way
\begin{equation}
\widehat{a}_{\mathbf{k}}\left\vert 0\right\rangle =\widehat{b}_{n\mathbf{k}%
}\left\vert 0\right\rangle =0,n=1,2,3, \label{vacuum}%
\end{equation}
for all $\mathbf{k}$. Since the classical fluid is invariant under spacetime
translations, it is natural to impose the invariance of the vacuum under
translations
\begin{equation}
\widehat{p^{\mu}}\left\vert 0\right\rangle =p^{\mu}\left\vert 0\right\rangle .
\label{inv-tr-vac}%
\end{equation}
The physical states are obtained by acting with the creation operators on the
vacuum state. For example, the one-particle excitations of the quantum fluid
potentials are described by the following states
\begin{equation}
\left\vert \mathbf{k}\right\rangle _{\theta}=\widehat{a}_{\mathbf{k}}^{\dag
}\left\vert 0\right\rangle ~~,~~\left\vert k_{n}\right\rangle _{\mathbf{A}%
}=\widehat{b}_{n\mathbf{k}}^{\dag}\left\vert 0\right\rangle ,n=1,2,3.
\label{one-part-excit}%
\end{equation}

In this way, the canonical quantization can be carried out straightforwardly
to the relativistic fluid models described by the hamiltonian
(\ref{mod-hamilton}). However, there are some differences from the general
field quantization due to the quantum equation of motion
(\ref{quantum-eq-motion}) that should be satisfied operatorially on all
quantum states, and to some arbitrariety in defining the quantum structure of
the vector potential $\widehat{\mathbf{A}.}$ Indeed, the equation
(\ref{quantum-eq-motion}) takes the following form in terms of creation and
annihilation operators
\begin{equation}
\left(  -\frac{\omega_{\mathbf{k}}^{2}}{c^{2}}+\mathbf{k}^{2}\right)  \left(
\widehat{a}_{\mathbf{k}}+\widehat{a}_{-\mathbf{k}}^{\dag}\right)  \left\vert
\psi\right\rangle =i\mathbf{k}\cdot\left(  \widehat{\mathbf{b}}_{\mathbf{k}%
}+\widehat{\mathbf{b}}_{\mathbf{k}}^{\dag}\right)  \left\vert \psi
\right\rangle .\label{quant-eq-a-daggera}%
\end{equation}
Note that for $\mathbf{k}=0$ either $\omega_{\mathbf{0}}=0$ or $\left(
\widehat{a}_{\mathbf{0}}+\widehat{a}_{\mathbf{0}}^{\dag}\right)  \left\vert
\psi\right\rangle =0$. If $\left\vert \psi\right\rangle =\left\vert
0\right\rangle $ then (\ref{quant-eq-a-daggera}) is equivalent to the
following relation
\begin{equation}
\left(  -\frac{\omega_{\mathbf{k}}^{2}}{c^{2}}+\mathbf{k}^{2}\right)
\left\vert \mathbf{k}\right\rangle _{\theta}=i\sum_{n=1}^{3}k_{n}\left\vert
k_{n}\right\rangle _{\mathbf{A}}.\label{vac-rel}%
\end{equation}
That shows that for $\mathbf{k}\neq0$ a different normalization of
$\widehat{b}_{n\mathbf{k}}^{\dag}\left\vert 0\right\rangle $ can be
considered
\begin{equation}
\widehat{b}_{n\mathbf{k}}^{\dag}\left\vert 0\right\rangle =-ik_{n}\left(
\frac{{\left\vert \mathbf{k}\right\vert ^{2}c^{2}-}\omega_{\mathbf{k}}^{2}%
}{{\left\vert \mathbf{k}\right\vert ^{2}c^{2}}}\right)  \left\vert
\mathbf{k}\right\rangle _{\mathbf{A}},\label{new-norm}%
\end{equation}
where $\left\vert \mathbf{k}\right\rangle _{\mathbf{A}}=\sum_{n=1}%
^{3}\mathbf{e}_{n}\left\vert k_{n}\right\rangle _{\mathbf{A}}$ and
$\{\mathbf{e}_{n}\}$, $n=1,2,3$ are orthogonal unit vectors on the spacelike
surface. The equations (\ref{vac-rel}) with the normalization (\ref{new-norm})
is identically satisfied by the dispersion relation of the massless scalar
field $\omega_{\mathbf{k}}^{2}=\left\vert \mathbf{k}\right\vert ^{2}c^{2}$.

In the above quantization, the unpropagating complex fluid potentials $z$ and
$\overline{z}$ and the K\"{a}hler potential field $K(z,\overline{z})$ were not
quantized directly, but rather under the form of the vector potential
$\mathbf{A}$. Therefore, in principle one could also consider the possibility
of the semiclassical quantized fluid in which the field $\mathbf{A}$ is a
classical vector potential and the dynamical canonical pair $(\theta
,\pi_{\theta})$ is quantized. However, by repeating the steps performed above
in the canonical quantization formalism, but now with $\hat{{\mathbf{A}}%
}=\mathbf{A}\hat{\mathbf{1}}$, where $\widehat{\mathbf{1}}$ is the identity
operator, we arrive at the following relation among the one-particle
excitations and the vacuum state
\begin{equation}
\left\vert \mathbf{k}\right\rangle _{\theta}=i\left(  \mathbf{k}^{2}%
-\frac{\omega^{2}}{c^{2}}\right)  \mathbf{k}\cdot\mathbf{b}_{\mathbf{k}%
}\left\vert 0\right\rangle _{\theta}. \label{semiclass}%
\end{equation}
That shows that if the Fourier coefficients $b_{n\mathbf{k}}$ are classical
functions on $z$ and $\overline{z}$, then the Fock space is highly degenerate
with all one-particle states proportional to the vacuum. Thus, we conclude
that if the relativistic fluid model of the type discussed here is to be
treated as a quantum system, the vector potential $\mathbf{A}$ should be
quantized as before in order to avoid the infinite vacuum degeneracy.

\section{Quantum Topological Charge}

In this section we are going to construct the quantum topological charge
(quantum linking number) operator which is the quantum counterpart of the
classical linking number, and to discuss the properties of the Fock space
states with respect to it.

The starting point is the classical topological linking number $\omega$
defined in the relation (\ref{classical-topological-number}). After some
algebraic manipulations, it can be put in the following form
\begin{equation}
\omega=\int d^{3}x\varepsilon^{lmn}\left(  \partial_{l}\theta\partial
_{m}\partial_{n}\theta+2\partial_{l}\theta\partial_{m}A_{n}+2A_{l}\partial
_{m}\partial_{n}\theta+4A_{l}\partial_{m}A_{n}\right)  .\label{t-charge}%
\end{equation}
Note that the first and the third terms from the above relation vanish due to
presence of the totally antisymmetric tensor in the integrand, while the
second term vanishes up to a total derivative. Let us define the quantum
topological charge operator $\widehat{\omega}$ by interpreting the fields in
(\ref{t-charge}) as quantum operators and taking the normal ordering of the
creation and annihilation operators in the mode expansion of the quantum
fields inside the integral. The ordering prescription is necessary due to the
presence of potentially divergent terms, e.g. $\delta^{3}\left(  0\right)
k_{l}k_{m}\varepsilon^{lmn}$, which for arbitrary value of $k_{n}$ are zero
and for $\left\vert k\right\vert \longrightarrow0$ have an undetermined limit.
They are a consequence of the commutation relation (\ref{comm-1}) among the
oscillator operators of $\widehat{\theta}$ field. By using the field expansion
relations (\ref{theta-fin}), (\ref{mom-theta-fin}) and (\ref{A-fin}) inside
the relation (\ref{t-charge}), the quantum topological charge operator can be
decomposed in the following sum
\begin{equation}
\colon\widehat{\omega}(x^{0})\colon=2\left(  2\pi\right)  ^{3}\left(
\colon\widehat{\omega}_{2}(x^{0})\colon+2i\colon\widehat{\omega}_{4}%
(x^{0})\colon\right)  ,\label{quantum-t-charge}%
\end{equation}
The operators $\colon\widehat{\omega}_{a}(x^{0})\colon$, $a=2,4$ from the r.
h. s. of the above relation represent the operatorial counterpart of the the
second term, kept due to its total derivative contribution, and the last term
from the relation (\ref{t-charge}), respectively,  in that order. After some
calculations, one arrives at the following explicit form of the operators
$\colon\widehat{\omega}_{a}(x^{0})\colon$ written in terms of creation and
annihilation operators
\begin{align}
\colon\widehat{\omega}_{2}(x^{0})\colon &  =\int d^{3}kN_{\mathbf{k}}%
^{2}\varepsilon^{lmn}\left[  \widehat{b}_{n\mathbf{k}}^{\dag}\widehat
{a}_{\mathbf{k}}+\widehat{a}_{\mathbf{k}}^{\dag}\widehat{b}_{n\mathbf{k}%
}+\widehat{a}_{\mathbf{k}}\widehat{b}_{n\mathbf{k}}e^{-\frac{2i\omega
_{\mathbf{k}}}{c}x^{0}}+\widehat{a^{\dag}}_{\mathbf{k}}\widehat{\mathbf{b}%
^{\dag}}_{n\mathbf{k}}e^{\frac{2i\omega_{\mathbf{k}}}{c}x^{0}}\right]
k_{l}k_{m},\label{omega2}\\
\colon\widehat{\omega}_{4}(x^{0})\colon &  =\int d^{3}kN_{\mathbf{k}}%
^{2}\varepsilon^{lmn}\left[  \widehat{b^{\dag}}_{l\mathbf{k}}\widehat
{b}_{n\mathbf{k}}-\widehat{b^{\dag}}_{n\mathbf{k}}\widehat{b}_{l\mathbf{k}%
}-\widehat{b}_{l\mathbf{k}}\widehat{b}_{n\mathbf{k}}e^{-\frac{2i\omega
_{\mathbf{k}}}{c}x^{0}}+\widehat{b}_{l\mathbf{k}}^{\dag}\widehat
{b}_{n\mathbf{k}}^{\dag}e^{\frac{2i\omega_{\mathbf{k}}}{c}x^{0}}\right]
k_{m}.\label{omega4}%
\end{align}
Since the components of the momentum $k_{n}$ commute with any operator, one
can see that $\colon\widehat{\omega}_{2}\colon$ vanishes due to the presence
of the totally antisymmetric tensor. The same symmetry arguments do not apply
to $\colon\widehat{\omega}_{4}\colon$ which is not zero. However, the time
dependent terms vanish in $\colon\widehat{\omega}_{4}\colon$. Indeed, by using
the commutation relations (\ref{comm-1}) in the terms $\varepsilon
^{lmn}\widehat{b}_{l\mathbf{k}}\widehat{b}_{n\mathbf{k}}k_{m}$ and
$\varepsilon^{lmn}\widehat{b}_{l\mathbf{k}}^{\dag}\widehat{b}_{n\mathbf{k}%
}^{\dag}k_{m}$, one obtains vanishing operator coefficients for each component
$k_{n}$. Therefore, the topological charge operator of the quantum
relativistic fluid from the class of models discussed here has the form
\begin{equation}
\colon\widehat{\omega}\colon=2i\lambda\hbar c\int\frac{d^{3}k}{\omega
_{\mathbf{k}}}\varepsilon^{lmn}k_{m}\left(  \widehat{b^{\dag}}_{l\mathbf{k}%
}\widehat{b}_{n\mathbf{k}}-\widehat{b^{\dag}}_{n\mathbf{k}}\widehat
{b}_{l\mathbf{k}}\right)  ,\label{final-top-charge-op}%
\end{equation}
where the dependence on  $x^{0}$ has been dropped from $\colon\widehat{\omega
}\colon$. Remark that the quantum topological number operator is parametrized
by the parameter $\lambda$ linearly and that it is expressed in $\hbar c$
units. Its dependence on the K\"{a}hler and complex potentials is hidden in
the integrand and has a simple form due to the quantization of $\mathbf{A}$
field. The fields $\widehat{\theta}$ and $\widehat{\pi}_{\theta}$ do not
contribute to the topological charge operator that is time independent.

In order to see how $\colon\widehat{\omega}\colon$ acts on the Fock space, we
consider a general state of fixed momentum vector $\mathbf{k}^{\prime}=\left(
k_{1}^{^{\prime}},k_{2}^{^{\prime}},k_{3}^{^{\prime}}\right)  $ of the form
\begin{equation}
\left\vert \Psi\left(  k_{1}^{^{\prime}},k_{2}^{^{\prime}},k_{3}^{^{\prime}%
}\right)  \right\rangle =\left\vert n_{1}\left(  k_{1}^{^{\prime}}\right)
,n_{2}\left(  k_{2}^{^{\prime}}\right)  ,n_{3}\left(  k_{3}^{^{\prime}%
}\right)  \right\rangle =%
{\displaystyle\bigotimes\limits_{q=1}^{3}}
\left\vert n_{q}\left(  k_{q}^{^{\prime}}\right)  \right\rangle
.\label{general-states-1}%
\end{equation}
Here, $n_{q}\left(  k_{q}^{^{\prime}}\right)  $ denotes the number $n_{q}$ of
excitations of $b_{q}$ type which have the value of momentum $k_{q}^{^{\prime
}}$ in the $q=1,2,3$ direction. The last equality in (\ref{general-states-1})
shows that the state factorizes as a tensor product of number operator
eigenstates in the corresponding directions \footnote{The dependence of the
state given in the relation (\ref{general-states-1}) on the excitations of the
scalar potential $\widehat{\theta}$ does not affect the calculation of the
topological charge since $\widehat{\theta}$ does not contribute to the
topological number operator.}. Also, we consider that the number of
excitations in each direction is fixed in the state (\ref{general-states-1})
and that the number operator eigenstates are ortogonal and normalized to
unity
\begin{equation}
\left\langle n_{l}\left(  k_{l}\right)  |m_{s}\left(  k_{s}^{^{\prime}%
}\right)  \right\rangle =\delta_{l,s}\delta_{n_{l},m_{s}}\delta\left(
k_{l}-k_{s}^{^{\prime}}\right)  .\label{normalization-oscillators}%
\end{equation}
It is easy to see that the expectation value of $\colon\widehat{\omega}\colon$
is zero in the states of the form (\ref{general-states-1}) because of the
fixed number of excitations in each direction one starts with. The vacuum
state is of the form (\ref{general-states-1}) and thus one concludes that the
vacuum has zero linking number as expected.

Since the fixed number of excitations in the state (\ref{general-states-1})
together with the ortogonality relation (\ref{normalization-oscillators}) make
the topological charge number vanish, one could take instead the more general
linear combination of states that have the following form%
\begin{equation}
\left\vert \Psi\right\rangle =%
{\displaystyle\int}
\frac{d^{3}k^{\prime}}{(2\pi)^{3}}\sum_{n_{1},n_{2},n_{3}=1}^{3}C_{n_{1}%
,n_{2},n_{3}}\left(  k_{1}^{^{\prime}},k_{2}^{^{\prime}},k_{3}^{^{\prime}%
}\right)  \left\vert n_{1}\left(  k_{1}^{^{\prime}}\right)  ,n_{2}\left(
k_{2}^{^{\prime}}\right)  ,n_{3}\left(  k_{3}^{^{\prime}}\right)
\right\rangle ,\label{general-states-2}%
\end{equation}
where the arbitrary complex coefficient functions $C_{n_{1},n_{2},n_{3}%
}\left(  k_{1}^{^{\prime}},k_{2}^{^{\prime}},k_{3}^{^{\prime}}\right)  $ can
be normalized to make the integral covariant if necessary. The states from the
above relations are entangled states of tensored number operators eigenstates.
One can show that the expectation value $\left\langle \colon\widehat{\omega
}\colon\right\rangle $ in the states of the form (\ref{general-states-2}) is
given by the following relation
\begin{align}
\left\langle \colon\widehat{\omega}\colon\right\rangle  &  =4i\lambda\hbar
c\int\frac{d^{3}k}{\omega_{\mathbf{k}}}\sum_{n_{1},n_{2},n_{3}=1}^{3}%
C_{n_{1},n_{2},n_{3}}\left(  k_{1},k_{2},k_{3}\right)  \nonumber\\
&  \times\left\{  k_{1}\left[  \sqrt{n_{2}\left(  n_{3}+1\right)  }%
\overline{C}_{n_{1},n_{2-1},n_{3+1}}\left(  k_{1},k_{2},k_{3}\right)
-\sqrt{n_{3}\left(  n_{2}+1\right)  }\overline{C}_{n_{1},n_{2+1},n_{3-1}%
}\left(  k_{1},k_{2},k_{3}\right)  \right]  \right.  \nonumber\\
&  +k_{2}\left[  \sqrt{n_{1}\left(  n_{3}+1\right)  }\overline{C}%
_{n_{1}+1,n_{2},n_{3-1}}\left(  k_{1},k_{2},k_{3}\right)  -\sqrt{n_{1}\left(
n_{3}+1\right)  }\overline{C}_{n_{1-1},n_{2},n_{3+1}}\left(  k_{1},k_{2}%
,k_{3}\right)  \right]  \nonumber\\
&  +\left.  k_{3}\left[  \sqrt{n_{1}\left(  n_{2}+1\right)  }\overline
{C}_{n_{1}-1,n_{2+1},n_{3}}\left(  k_{1},k_{2},k_{3}\right)  -\sqrt
{n_{2}\left(  n_{1}+1\right)  }\overline{C}_{n_{1+1},n_{2-1},n_{3}}\left(
k_{1},k_{2},k_{3}\right)  \right]  \right\}  ,\label{quantum-top-number}%
\end{align}
Some comments are in order now. Firstly, note that the quantum topological
number given in the relation (\ref{quantum-top-number}) is different from zero
if the integral does not vanish. This condition should be satisfied by a large
set of arbitrary coefficients $C_{n_{1},n_{2},n_{3}}\left(  k_{1}^{^{\prime}%
},k_{2}^{^{\prime}},k_{3}^{^{\prime}}\right)  $ which thus define the states
with non-vanishing quantum topological linking number. Secondly, one can see
from the classical state equations of the model (\ref{toy-model}) and the
relation (\ref{quantum-top-number}), that there is a large number of quantum
states in which the quantum linking number is infinite due to the infinite
value of the classical limit of the pressure density or the energy density, i.
e. for $p\rightarrow\infty$ or $\varepsilon\rightarrow\infty$, respectively.
On the other hand, for $p\rightarrow0$ or $\varepsilon\rightarrow0$, the
topological number is zero, unless the integral and sum in the r. h. s. of
(\ref{quantum-top-number}) diverge in the corresponding state. Thirdly, note
that in general either the quantum linking number is zero or it is time
independent as expected. This result expresses the conservation of the
expectation values of the linking number operator.

\section{Conclussions}

In this paper, we have investigated the quantization of the relativistic fluid
on K\"{a}hler manifolds. The class of models considered here are parametrized
by an arbitrary K\"{a}hler potential depending on two complex fluid potentials
$z$ and $\overline{z}$ and a real parameter $\lambda$. That type of models
represents a subset of a larger set parametrized by $\{K(z,\overline
{z}),f(\rho)\}$ that was firstly proposed in \cite{nhg1}. Due to the arbitrary
dependence of the lagrangian on $\rho=\sqrt{\pi_{\theta}^{2}-\mathbf{j}^{2}}$,
the full set of relativistic fluid models is not suitable for quantization. As
was shown in \cite{nhg1}, the degrees of freedom of the full set
$\{K(z,\overline{z}),f(\rho)\}$ are constrained by second class constraints.
As a first result, we have obtained the physical degrees of freedom of the
fluid described by $\{K(z,\overline{z}),f(\rho)\}$ by applying the symplectic
projector method. We have concluded that the classical theory on the physical
surface displays the same topological charges as the original theory
\cite{nhg1} since the symplectic projector does not include the conserved
axial currents. Our results for the full set of classical relativistic fluids
agree with the ones presented in the literature which revalidates the
applicability of the symplectic projector method to second class constraints.

The main result of this paper is the quantization of the smaller set of models
$\{K(z,\overline{z}),\lambda\}$. We have obtained the quantum theory by
applying the canonical quantization methods to the pair of fields $(\theta
,\pi_{\theta})$ as well as to the vector field $\mathbf{A}(K,z,\overline{z})$
which encodes the information about the K\"{a}hler and the complex fluid
potentials, respectively, and we have constructed the Fock space of the
relativistic fluid and the one-particle excitations of the relativistic
potentials. Also, we have discussed the semiclassical quantization of the
relativistic fluid in which the potential $\mathbf{A}(K,z,\overline{z})$ is a
classical field. By analysing the one-particle spectrum, it has been shown
that the vacuum of the semiclassical theory is infinitely degenerate. From
that, one concludes that $\widehat{\mathbf{A}}$ should be treated as a quantum field.

The second important result of the present paper is the construction of the
quantum linking number operator $\widehat{\omega}$ which was defined by taking
the normal ordered field products in the r. h. s. of the relation
(\ref{t-charge}). The operator $\widehat{\omega}$ is time independent and it
is determined only by the K\"{a}hler and the complex potentials, with no
contribution from the real potential. We have shown that the vacuum of the
quantum relativistic fluid has vanishing linking number, as well as the states
formed by taking the tensor product of number operators eigenstates of
$\widehat{\mathbf{A}}$ field. However, there are entangled number operator
eigenstates with non-vanishing linking number.

As a final comment, we note that the classical topological number can be
expressed as a surface term as in the relation
(\ref{classical-topological-number}). It would be certainly interesting to
compare the approach presented in this paper with a different quantization
method and to attempt a proper treatment of the boundaries in the quantum
theory. That analysis and the application of the present method to the
supersymmetric fluid \cite{jpp2} in the K\"{a}hler parametrization will
hopefully be discussed elsewhere.

\noindent\textbf{Acknowledgements} We thank to H. Belich for useful
discussions. I. V. V. would like to acknowledge J. C. Fabris for hospitality
at DEFIS - UFES. L. H. and I. V. V. would like to thank to J. A.
Helay\"{e}l-Neto, S. A. Dias and A. M. O. de Almeida for hospitality at
LAFEX-CBPF where parts of this work were accomplished. Finally, we would like
to acknowledge an anonymous referee for constructive discussions.


\end{document}